\documentclass[aps,prl, twocolumn, superscriptaddress, longbibliography]{revtex4-1}
\usepackage{amsmath}
\usepackage{amssymb}
\usepackage{graphicx}
\usepackage[caption=false, position=top, singlelinecheck=off, justification=raggedright]{subfig}
\usepackage{color}
\usepackage{pst-node}
\usepackage[unicode]{hyperref}
\hypersetup{unicode=true, colorlinks=true, linkcolor=blue, citecolor=blue}

\begin{document}

\title{Observation of a dissipative time crystal}

\author{Hans Ke{\ss}ler}
\affiliation{Zentrum f\"ur Optische Quantentechnologien and Institut f\"ur Laser-Physik, 
Universit\"at Hamburg, 22761 Hamburg, Germany}

\author{Phatthamon Kongkhambut}
\affiliation{Zentrum f\"ur Optische Quantentechnologien and Institut f\"ur Laser-Physik, Universit\"at Hamburg, 22761 Hamburg, Germany}

\author{Christoph Georges}
\affiliation{Zentrum f\"ur Optische Quantentechnologien and Institut f\"ur Laser-Physik, Universit\"at Hamburg, 22761 Hamburg, Germany}

\author{Ludwig Mathey}
\affiliation{Zentrum f\"ur Optische Quantentechnologien and Institut f\"ur Laser-Physik, 
Universit\"at Hamburg, 22761 Hamburg, Germany}
\affiliation{The Hamburg Center for Ultrafast Imaging, Luruper Chaussee 149, 22761 Hamburg, Germany}

\author{Jayson G. Cosme}
\affiliation{National Institute of Physics, University of the Philippines, Diliman, Quezon City 1101, Philippines}

\author{Andreas Hemmerich}
\affiliation{Zentrum f\"ur Optische Quantentechnologien and Institut f\"ur Laser-Physik, Universit\"at Hamburg, 22761 Hamburg, Germany}
\affiliation{The Hamburg Center for Ultrafast Imaging, Luruper Chaussee 149, 22761 Hamburg, Germany}

\date{\today}

\begin{abstract}
We present the first experimental realisation of a time crystal stabilized by dissipation. The central signature in our implementation in a driven open atom-cavity system is a period doubled switching between distinct chequerboard density wave patterns, induced by the interplay between controlled cavity-dissipation, cavity-mediated interactions and external driving. We demonstrate the robustness of this dynamical phase against system parameter changes and temporal perturbations of the driving.
\end{abstract}

\pacs{PACS numbers: 03.75.-b, 42.50.Gy, 42.60.Lh, 34.50.-s}

\maketitle
Phase transitions of matter can be associated with the spontaneous breaking of a symmetry. For crystallization, this broken symmetry is the spatial translation symmetry, as the atoms spontaneously localize in a periodic arrangement. In analogy to spatial crystals, the spontaneous breaking of temporal translation symmetry can result in the formation of  time crystals. Following its initial proposal\cite{Wilczek12, Shapere2012}, the possibility of time crystals in the ground state of equilibrium many-body systems was ruled out for fundamental reasons \cite{Bruno2013, Watanabe2015}. This development led to a paradigm shift, directing the search for time crystals towards genuine nonequilibrium scenarios \cite{Zhang2017, Choi2017, Rovny2018, Sacha2018, else20, osullivan20, lazarides20}. In particular, the no-go theorem \cite{Bruno2013, Watanabe2015} can be circumvented by periodic driving, which imposes a discrete time translation symmetry on the system. Floquet or discrete time crystals emerge, when discrete time translation symmetry is spontaneously broken, which manifests as a subharmonic response of an observable \cite{Sacha2015, Else2016, Yao2017, khemani16}. Previous experimental studies have focused on driven closed quantum systems with long-lived time crystalline response enabled by many-body mechanisms, which impede excessive heating \cite{Zhang2017, Choi2017, Rovny2018, Autti2018, smits18}. However, as proposed by theoretical work \cite{chitra15, gong18, zhu19, Buca2019, Yao2020}, dissipation and fluctuations, induced via controlled coupling to a suitable environment can also serve as a source for stabilization of time-crystal dynamics.

\begin{figure}[ht!]
\centering
\includegraphics[width=0.92\columnwidth]{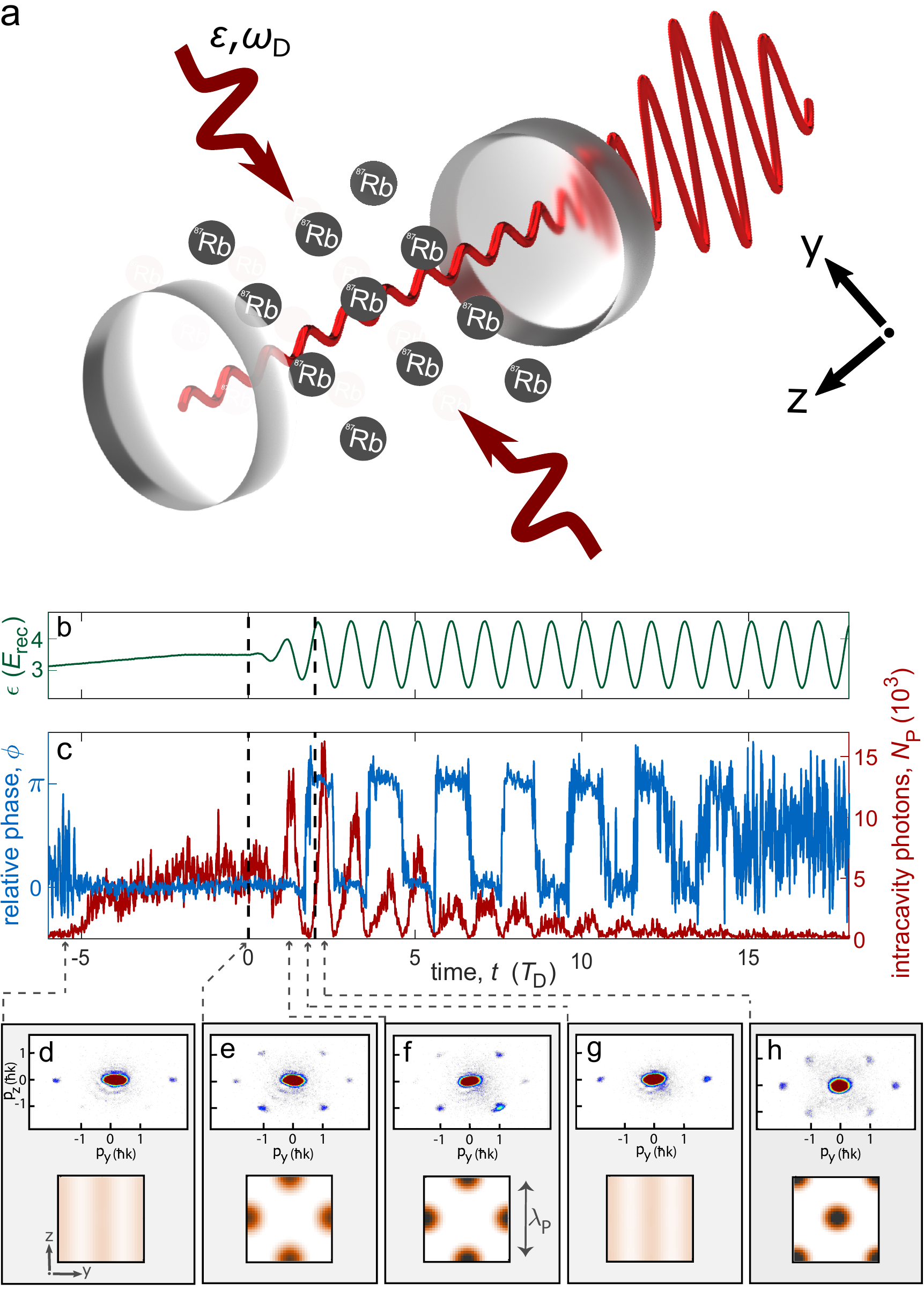}
\caption{{Dissipative time crystal.} {(a)} Schematic diagram of the transversely pumped atom-cavity system.  {(b)} Time sequence for the pump with modulation strength $f_\mathrm{0}=$~0.3 and modulation period $T_\mathrm{D} = 0.25\,$ms. In the time interval delimited by dashed lines, $f_\mathrm{0}$ is linearly ramped from zero to its desired value.  {(c)} The corresponding response of the intracavity photon number $N_\mathrm{P}$ (red) and the relative phase $\phi$ between the pump and the cavity light field (blue).  {(d)-(h)} Top panels: momentum distributions measured at instances of time marked by dashed arrows in {(c)}.  Bottom panels: corresponding mean-field results for the single-particle density distribution, which shows periodic switching between even and odd DWs at twice the driving period.}
\label{fig:1} 
\end{figure}

\begin{figure*}[!t]
\centering
\includegraphics[width=2\columnwidth]{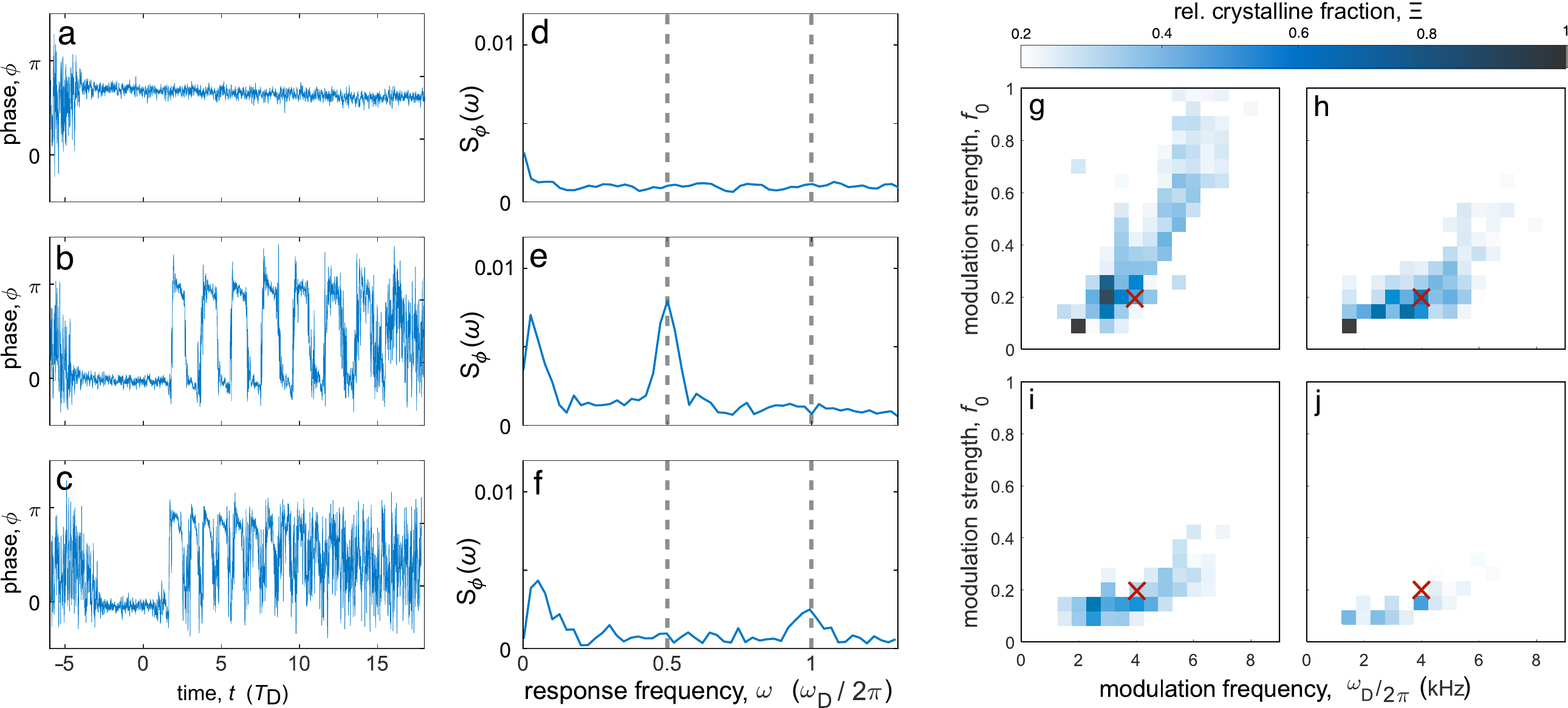}
\caption{Dynamical regimes.  Dynamics of the relative phase $\phi$ for ({a}) $f_\mathrm{0}= 0.05$ ,   ({b}) $f_\mathrm{0}=0.25$, and ({c}) $f_\mathrm{0}=0.95$ with fixed $\omega_\mathrm{D}=2\pi \times 4~\mathrm{kHz}$.  {(d)-(f)} Corresponding Fourier spectra of the dynamics in ({a})-({c}). As the modulation strength is increased, the system transforms from a DW to a DTC phase. Strong modulation leads to heating and chaotic behaviour.  (g)-(j) Dynamical phase diagram showing the relative crystalline fraction $\Xi$ as a function of the modulation frequency $\omega_\mathrm{D}$ and strength $f_\mathrm{0}$ for ({g}) clean modulation,  ({h}) weak noise strength $n=9.6$,  ({i}) intermediate noise strength $n=15.9$, and ({j}) large noise strength $n=22.3$. The diagram is constructed by dividing the parameter space into $18 \times 18$ plaquettes and within each averaging over multiple experimental runs (at least four realizations). Red crosses in {(g)-(j)} mark the modulation parameters used in Figs.~\ref{fig:3}(a)-\ref{fig:3}(d). Increasingly large noise strengths shrink the area in the phase diagram where a stable DTC phase prevails.}
\label{fig:2} 
\end{figure*}

Here, we report the experimental realisation of a dissipative time crystal (DTC) phase in an atom-cavity platform \cite{Ritsch2013}. This is inspired by a recent theoretical proposal for a time crystal stabilised through an interplay between interaction and dissipation in the open Dicke model, arising when the light-matter coupling is periodically modulated \cite{chitra15,gong18,zhu19}. The defining feature of this paradigmatic DTC is a subharmonic response, where the system periodically switches between pairs of $\mathbb{Z}_2$ symmetry broken superradiant states.

A Bose-Einstein condensate (BEC) of $^{87}$Rb atoms is prepared inside a high-finesse optical cavity pumped by a retro-reflected laser beam at wavelength $ \lambda_\mathrm{P} = 803\,$nm, aligned perpendicular to the cavity axis, as depicted in Fig.~\ref{fig:1}(a). The atom-cavity system operates in the recoil-resolved regime, where the cavity field and the atomic distribution evolve at a similar timescale leading to a retarded infinite-range cavity-mediated interaction between the atoms\cite{Klinder2016}. Above a critical value of the pump strength $\epsilon$, the system undergoes a self-organisation transition from a BEC phase to a density wave (DW) phase, which emulates the superradiant phase transition in the open Dicke model \cite{Baumann:2010js,Klinder:0fv}. In a spontaneous $\mathbb{Z}_2$ symmetry breaking process, an intracavity optical lattice arises, which traps the atoms either in the black or the white squares of a chequerboard pattern, denoted as odd and even DW. 

An effective driving of the light-matter coupling can be realized by modulating the pump strength. Off-resonant driving of the pump strength at a frequency $\omega_\mathrm{D}$ notably exceeding the recoil frequency $\omega_{\mathrm{rec}} \equiv \hbar k^2/(2 m) = 2\pi \times 3.55\,$kHz, with $k \equiv 2\pi / \lambda_\mathrm{P}$ and the atomic mass $m$, leads to a dynamical renormalisation of the phase boundary between the BEC and DW phases \cite{Cosme2018,Georges2018}. On the other hand, a period doubling response characterised by periodic switching between the odd and even DWs has been predicted for modulating only slightly above the recoil frequency \cite{chitra15,molignini18,Cosme2019}. This phase, originally addressed as dynamical normal phase \cite{chitra15}, shows subharmonic oscillations between the two $\mathbb{Z}_2$ symmetry broken even and odd DW states and is closely related to the DTC phase proposed in the open Dicke model \cite{gong18}. In the thermodynamic limit, $N \to \infty$, the Dicke model can be transformed into a parametrically driven coupled oscillator system with two polaritonic states. Driving at twice the lower polariton frequency leads to an instability, which gives rise to a period-doubling response in the full atom-cavity model (cf. \cite{supp}). In the following, we describe the experimental realization of a DTC in our atom-cavity system and analyse its properties as a time crystal.

\begin{figure*}[!t]
\centering
\includegraphics[width=2\columnwidth]{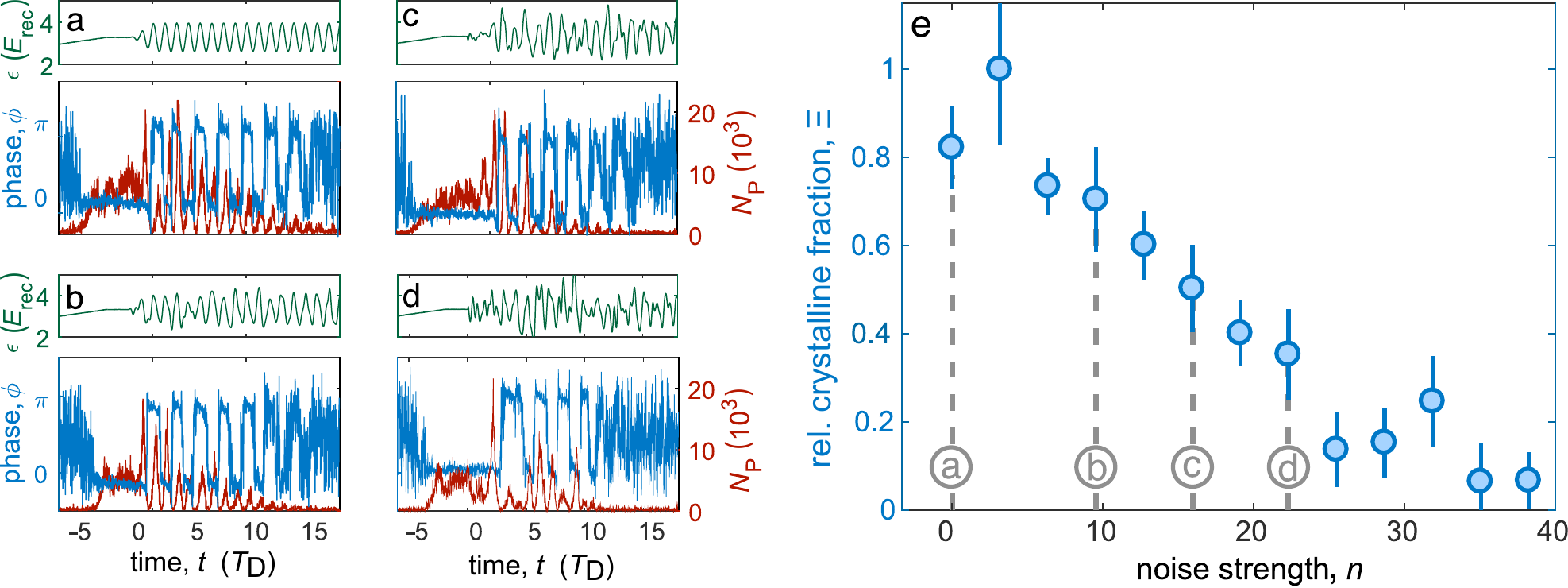}
\caption{{Robustness of subharmonic response.} {(a)-(d)} Single-shot experimental runs for the noise strengths applied 
in Figs.~\ref{fig:2}(g)-\ref{fig:2}(j) and for values of $\omega_{\mathrm{D}}$ and $f_\mathrm{0}$ according to the red crosses in these figures. Top panels: single-shot protocols for the pump strength. Bottom panel: corresponding time evolution of the relative phase $\phi$ (blue trace) and intracavity photon number $N_{\mathrm{P}}$ (red trace).  {(e)} Dependence of the relative crystalline fraction $\Xi$ on the noise strength averaged over 7 experimental runs with $f_\mathrm{0}=0.2$ and $\omega_\mathrm{D}=2\pi \times 4$~kHz. The gray dashed lines mark the noise strengths used in {(a)-(d)}.}
\label{fig:3} 
\end{figure*}

Each experimental sequence begins with preparing the atom-cavity system in the self-organized DW phase (see \cite{supp}). An example of a time sequence for the pump is shown in Fig.~\ref{fig:1}(b). For $t < -5\,T_\mathrm{D}$ the system is in the BEC phase. The intracavity photon number $N_\mathrm{P}$ is zero and the observed momentum spectrum in the upper panel of Fig.~\ref{fig:1}(d) shows the BEC mode at zero momentum and two Bragg resonances at $\pm 2\hbar k$ along the $y$-direction, associated with the matter grating induced by the pump wave. This grating is illustrated in the lower panel of Fig.~\ref{fig:1}(d) by showing the single-particle density distribution obtained from a mean-field model (see \cite{supp}). The self-organisation transition into the DW phase is observed in Fig.~\ref{fig:1}(c) around $t \approx -5\,T_\mathrm{D}$, as evinced from a significant increase in the intracavity photon number $N_\mathrm{P}$ and the locking of the relative phase $\phi$ between the pump and cavity fields at a constant value $\phi \approx 0$. A momentum spectrum, characteristic for the DW phase, is shown in the upper panel of Fig.~\ref{fig:1}(e) for $t=0$. The occupation of the momentum modes $\{p_\mathrm{y},p_\mathrm{z}\}=\{\pm \hbar k, \pm \hbar k\}$ signals the formation of an intracavity chequerboard matter grating, as illustrated by the calculated single-particle density distribution, shown in the lower panel. The two possible energetically degenerate DW states can be distinguished by their associated values of the phase $\phi=0$ or $\phi=\pi$ for odd and even realizations, respectively \cite{Baumann:2011io}. We measure $N_\mathrm{P}$ and $\phi$ using a balanced heterodyne detection scheme \cite{pino11}. The probability for the occurrence of the odd and even DW configurations is found to be close to $50 \%$ (see \cite{supp}), which confirms that the discrete symmetry breaking in the chequerboard DW phase is well established in our system.

Upon preparation of the DW phase, in the time interval delimited by the vertical dashed lines, we linearly increase the modulation strength $f_\mathrm{0}$ in $500\,\mu$s from zero to its final value (see Fig.~\ref{fig:1}(b)). Subsequently, $f_\mathrm{0}$ is kept constant for $5\,$ms, such that the pump strength evolves according to $\epsilon=\epsilon_\mathrm{0} [ 1 + f_\mathrm{0} \sin(\omega_\mathrm{D}  t)]$. The dynamical response seen in Fig.~\ref{fig:1}(c) for positive $t$, presents the key observation of this work: the emergence of a DTC phase characterised by pulsating behaviour of the intracavity photon number $N_\mathrm{P}$ (red trace) and a period-doubling response of the relative phase $\phi$ (blue trace). The presence of intracavity photons highlights the many-body aspect of the DTC phase since they induce a retarded infinite-range interaction or all-to-all coupling between the atoms. The period-doubling dynamics arises in the relative phase $\phi$. As $\phi$ switches from zero to $\pi$ or vice-versa after one modulation cycle, the atomic ensemble self-organises from one type of chequerboard lattice (see Fig.~\ref{fig:1}(f)) to its symmetry-broken partner (see Fig.~\ref{fig:1}(h)). That is, the system requires two modulation cycles to return to its initial configuration. After half of a modulation period, the system crosses from the DW phase with significant occupation of the cavity mode, to the BEC phase, where the cavity is almost empty. This behaviour, corroborated by the momentum distribution in Fig.~\ref{fig:1}(g), is responsible for the pulsating intracavity photon number in Fig.~\ref{fig:1}(c) (red trace).

In Fig.~\ref{fig:2}(a)-\ref{fig:2}(f), we present the various dynamical regimes accessed by tuning the modulation strength. For weak modulation (see Figs.~\ref{fig:2}(a) and \ref{fig:2}(d)), the system stays in the DW phase and the relative phase remains locked to its value before the pump modulation. For intermediate modulation strength, the relative phase exhibits period-doubling dynamics (see Fig.~\ref{fig:2}(b)), resulting in a subharmonic peak at $\omega = \omega_\mathrm{D}/2$ in the Fourier spectrum in Fig.~\ref{fig:2}(e).  Increasing the modulation strength even further leads to chaotic dynamics dominated by heating and loss of spatiotemporal coherence (see Figs.~\ref{fig:2}(c) and \ref{fig:2}(f)). In contrast to the coherent switching observed in the DTC phase, the chaotic phase is characterised by intermittent dynamics of the relative phase, whereby the system appears to get stuck in one type of chequerboard pattern for two or more consecutive driving cycles (see Fig.~\ref{fig:2}(c)).

\begin{figure}[!htbp]
\centering
\includegraphics[width=1.0\columnwidth]{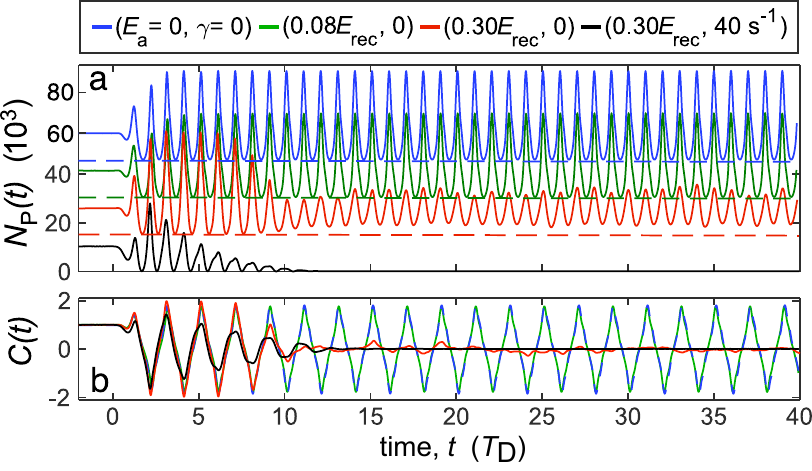}
\caption{{Short-range interaction and atom loss.} Numerical results from TWA for the dynamics of ({a}) the intracavity photon number $N_{\mathrm{P}}$ and ({b}) the non-equal time correlation $C$ of the photons for different values of the collisional interaction energy $E_a$ and the atom loss rate $\gamma$. To increase the quality of the presentation, the black, red, green, and blue traces in \textbf{a} are plotted with different offsets $0, 15, 30, 45 \times 10^3$, respectively. The modulation parameters are $\omega_\mathrm{D}=2\pi \times 4~\mathrm{kHz}$ and $f_\mathrm{0}=0.2$.}
\label{fig:4} 
\end{figure}

Next, we test the robustness of the DTC against variations of the system parameters and temporal perturbations. To this end, we calculate the relative crystalline fraction $\Xi$ \cite{Choi2017,Rovny2018}, defined by means of the amplitude of the subharmonic peak in the normalised Fourier spectrum $S_{\phi}(\omega)$ of the relative phase $\phi$ rescaled by its maximum, i.e., $\Xi=S_\mathrm{\phi}(\omega_\mathrm{D}/2)/S_\mathrm{max,\phi}$, where $S_\mathrm{max,\phi}$ is the maximum crystalline fraction measured in the parameter space spanned by $f_0\in[0,1]$ and $\omega_{\mathrm{D}}\in~2\pi\times[0,9]~\mathrm{kHz}$. Figure~\ref{fig:2}(g) displays the relative crystalline fraction for varying modulation parameters $\omega_\mathrm{D}$ and $f_\mathrm{0}$. We observe large relative crystalline fractions $\Xi >0.2$ for modulation frequencies $\omega_\mathrm{D}\in2\pi\times[2,8]$~kHz signalling a robust DTC order for a wide range of modulation parameters. Note that the overall shape of the relative crystalline fraction in Fig.~\ref{fig:2}(g) resembles the stability island of the DTC obtained from numerical simulations using a simple mean-field model (see \cite{supp} and Fig.~3 in \cite{supp}).

To explore the robustness of the DTC against temporal perturbations, we introduce a disorder in time by superimposing Gaussian white noise onto the signal of the pump strength. The noise strength is measured by $n \equiv \sum_{\omega=0}^{2\pi \times 50~\mathrm{kHz}} |\mathcal{E}_\mathrm{noisy}(\omega)|/\sum_{\omega=0}^{2\pi \times 50~\mathrm{kHz}} |\mathcal{E}_\mathrm{clean}(\omega)|$,  where $\mathcal{E}_\mathrm{noisy}$ ($\mathcal{E}_\mathrm{clean}$) is the Fourier spectrum of the pump in the presence (absence) of white noise. Figures~\ref{fig:2}(h)-\ref{fig:2}(j) show how the relative crystalline fraction changes with increasing noise strength. The area with clear DTC response, i.e., a large relative crystalline fraction, shrinks as the noise strength increases. Nevertheless, we still find a sizeable region, where a DTC phase exists, for relatively large noise strength (Fig.~\ref{fig:2}(j)). For a fixed set of modulation parameters marked by the red crosses in Figs.~\ref{fig:2}(g)-\ref{fig:2}(j), typical single-shot results for varying noise strengths are depicted in Figs.~\ref{fig:3}(a)-\ref{fig:3}(d). Note that even for a strongly distorted pump signal, as in Figs.~\ref{fig:3}(c) and \ref{fig:3}(d), the system still switches multiple times between the two sublattices before the intracavity photons disappear. The relative crystalline fraction at fixed modulation parameters decreases with increasing noise strength, as shown in Fig.~\ref{fig:3}(e). The small offset for large noise strength $n > 25$ is due to the background noise in the Fourier spectrum (see Fig.~\ref{fig:2}(f)). Our experimental findings suggest that the DTC in the modulated atom-cavity system is robust against fluctuations not only from the non-unitary dynamics of the dissipative cavity but also from temporal disorder added via driving.

Finally, we address the decay of the time-translation symmetry breaking response in the DTC phase, for example, seen in Fig.~\ref{fig:3}(a). The experimental lifetimes of time crystal implementations are generally finite due to a combination of technical limitations and undesired relaxation dynamics \cite{Zhang2017,Choi2017,Rovny2018,Autti2018,smits18}. In our experiment, the main cause for the decay of time-crystal dynamics can be attributed to two factors: (i) short-range collisional interaction and (ii) atom losses. In the case of the open Dicke model, the all-to-all coupling between the atoms makes it amenable to a mean-field description. In this theoretical limit, the mean-field solvability of the Dicke model provides the Dicke DTC with the necessary long-range spatiotemporal order and robustness such that it can persists to infinitely long times\cite{gong18, zhu19}. However, when mean-field breaking terms are present, the DTC may become unstable, leading to a decay of the symmetry breaking response \cite{zhu19}. In the atom-cavity system, the short-range interaction between the atoms competes with the collective coupling, induced by the cavity photons, and breaks the mean-field solvability of the model. To investigate the damping effects of short-range interaction and atom loss, we employ the truncated Wigner approximation (TWA). The transversely pumped atom-cavity system is thereby modelled by considering only the degrees of freedom along the pump ($y$ direction) and the cavity ($z$ direction) axes (see \cite{supp}). The short-range interaction is quantified in terms of the mean-field collisional interaction energy $E_{\mathrm{a}} = (U_{\mathrm{a}}/N_{\mathrm{a}})\int dy dz |\psi_0(y,z)|^4$, where $U_{\mathrm{a}}$ denotes the effective two-dimensional interaction strength (see \cite{supp}), $N_{\mathrm{a}}$ is the number of atoms and $\psi_0(y,z)$ is the wave function of the homogeneous BEC. We also include in our model a phenomenological atom loss channel described by $dN_{\mathrm{a}}/dt = -2\gamma \, N_{\mathrm{a}}$. We simulate the dynamics of the intracavity photon number $N_{\mathrm{P}}=\langle \hat{a}^{\dagger}\hat{a} \rangle$, where  $\hat{a}$ ($\hat{a}^{\dagger}$) is the bosonic operator that annihilates (creates) a photon in the single-mode cavity. To characterise temporal long-range order, we calculate the two-point temporal correlation function $C(t)=\mathrm{Re}[\langle a^{\dagger}(t)a(t_0) \rangle]/\langle a^{\dagger}(t_0)a(t_0)\rangle$ at $t_0=0$, the time before the modulation is switched on.

The resulting evolution of $N_{\mathrm{P}}(t)$ and $C(t)$ is studied in Fig.~\ref{fig:4} for different values of $E_{\mathrm{a}}$. First, we consider the dynamics in the absence of atom loss. For weak interaction strength $E_{\mathrm{a}}=0.08E_{\mathrm{rec}}$, $N_{\mathrm{P}}(t)$ and $C(t)$ in the green traces of Fig.~\ref{fig:4} are practically indistinguishable from the findings for $E_{\mathrm{a}}=0$ in the blue traces. However, stronger short-range interactions with $E_{\mathrm{a}}=0.30E_{\mathrm{rec}}$ lead to a metastable DTC, where the period-doubling oscillations in $C(t)$ decay after a few cycles, as seen in the red trace in Fig.~\ref{fig:4}(b). This translates to irregular dynamics of the corresponding intracavity photon number $N_{\mathrm{P}}(t)$ (red trace in Fig.~\ref{fig:4}(a)) in the long-time regime. Introducing an atom loss channel with $\gamma=40~s^{-1}$, which models the observed atom decay rate in the experiment, yields exponentially decaying behaviour as shown in the black traces in Fig.~\ref{fig:4}. This behaviour closely resembles the characteristic exponential decay of $N_{\mathrm{P}}$ in our experiment, such that the cavity is almost empty for $t/T_{\mathrm{D}} > 15$ (see Figs.~\ref{fig:3}(a)-\ref{fig:3}(d)). Atom loss leads to a trivial suppression of the atom-cavity coupling and hence of intracavity photons.  When the number of intracavity photons falls below the detection level, the relative phase $\phi$ becomes ill-defined leading to the fast and irregular oscillations of $\phi$ seen in Figs.~\ref{fig:3}(a)-\ref{fig:3}(d) for late times. Since we are operating close to the phase boundary between the DW and the normal phase the system is very sensitive to atom loss, which primarily limits the DTC lifetime in the experiment.

Our observations confirm the realization of a dissipative time crystal in an atom-cavity system with the defining feature of period-doubling dynamics. This quintessential DTC is robust against changes of the system parameters and temporal perturbations added to the drive, thereby fulfilling the robustness property of time crystals. Numerical results based on a simplified semiclassical model imply that short-range interaction and atom loss limits the lifetime of the DTC phase. 

\begin{acknowledgments}
This work is funded by the Deutsche Forschungsgemeinschaft (DFG, German Research Foundation) – SFB-925 – project 170620586 and the Cluster of Excellence “Advanced Imaging of Matter” (EXC 2056), Project No. 390715994.
\end{acknowledgments}

Note: During submission of this work, a subsequent example of dissipative time crystal was reported in an all-optical system \cite{taheri20}.

\nocite{Ritsch2013,Nagy:2008hk,Polkovnikov2010,Blakie2008,Carusotto2013,Cosme2019,Kessler2020}

\bibliography{references}

\end{document}


\title{Supplemental Material for\\Observation of a dissipative time crystal}

\author{Hans Ke{\ss}ler}
\affiliation{Zentrum f\"ur Optische Quantentechnologien and Institut f\"ur Laser-Physik, 
Universit\"at Hamburg, 22761 Hamburg, Germany}

\author{Phatthamon Kongkhambut}
\affiliation{Zentrum f\"ur Optische Quantentechnologien and Institut f\"ur Laser-Physik, Universit\"at Hamburg, 22761 Hamburg, Germany}

\author{Christoph Georges}
\affiliation{Zentrum f\"ur Optische Quantentechnologien and Institut f\"ur Laser-Physik, Universit\"at Hamburg, 22761 Hamburg, Germany}

\author{Ludwig Mathey}
\affiliation{Zentrum f\"ur Optische Quantentechnologien and Institut f\"ur Laser-Physik, 
Universit\"at Hamburg, 22761 Hamburg, Germany}
\affiliation{The Hamburg Center for Ultrafast Imaging, Luruper Chaussee 149, 22761 Hamburg, Germany}

\author{Jayson G. Cosme}
\affiliation{National Institute of Physics, University of the Philippines, Diliman, Quezon City 1101, Philippines}

\author{Andreas Hemmerich}
\affiliation{Zentrum f\"ur Optische Quantentechnologien and Institut f\"ur Laser-Physik, Universit\"at Hamburg, 22761 Hamburg, Germany}
\affiliation{The Hamburg Center for Ultrafast Imaging, Luruper Chaussee 149, 22761 Hamburg, Germany}

\maketitle

\section{Experimental details}
The experimental set-up, as sketched in Fig.~1a in the main text, is comprised of a magnetically trapped BEC of $N_\mathrm{a} = 65\times 10^3$  $^{87}$Rb atoms, dispersively coupled to a narrow-band high-finesse optical cavity. The cavity field has a decay rate of $\kappa=~2\pi \times$4.55~kHz, which is the same order of magnitude as the recoil frequency $\omega_\mathrm{rec} = E_\mathrm{rec}/\hbar  =~2\pi \times$3.55~kHz. The wavelength of the pump laser is $\lambda_\mathrm{P} = 803\,$nm, which is red detuned with respect to the relevant atomic transition of $^{87}$Rb at 795~nm. The maximum light shift per atom is $U_0 = -2\pi \times 0.36~\mathrm{Hz}$. We fix the effective detuning to $\delta_{\mathrm{eff}} \equiv \delta_{\mathrm{C}} - (1/2)N_\mathrm{a} U_0=-2 \pi \times 18.5~\mathrm{kHz}$, where $\delta_{\mathrm{C}} = \omega_{\mathrm{P}}-\omega_{\mathrm{C}}$ is the pump-cavity detuning. An experimental sequence starts by preparing the system in the self-organized density wave (DW) phase. This is achieved by linearly increasing the pump strength $\epsilon$ from zero to its final value $\epsilon_\mathrm{0}=3.3~E_\mathrm{rec}$ in 10~ms at a fixed pump-cavity detuning $\delta_\mathrm{eff}=-2\pi \times$18.5~kHz.

\begin{figure}[!htbp]
\centering
\includegraphics[width=0.5\columnwidth]{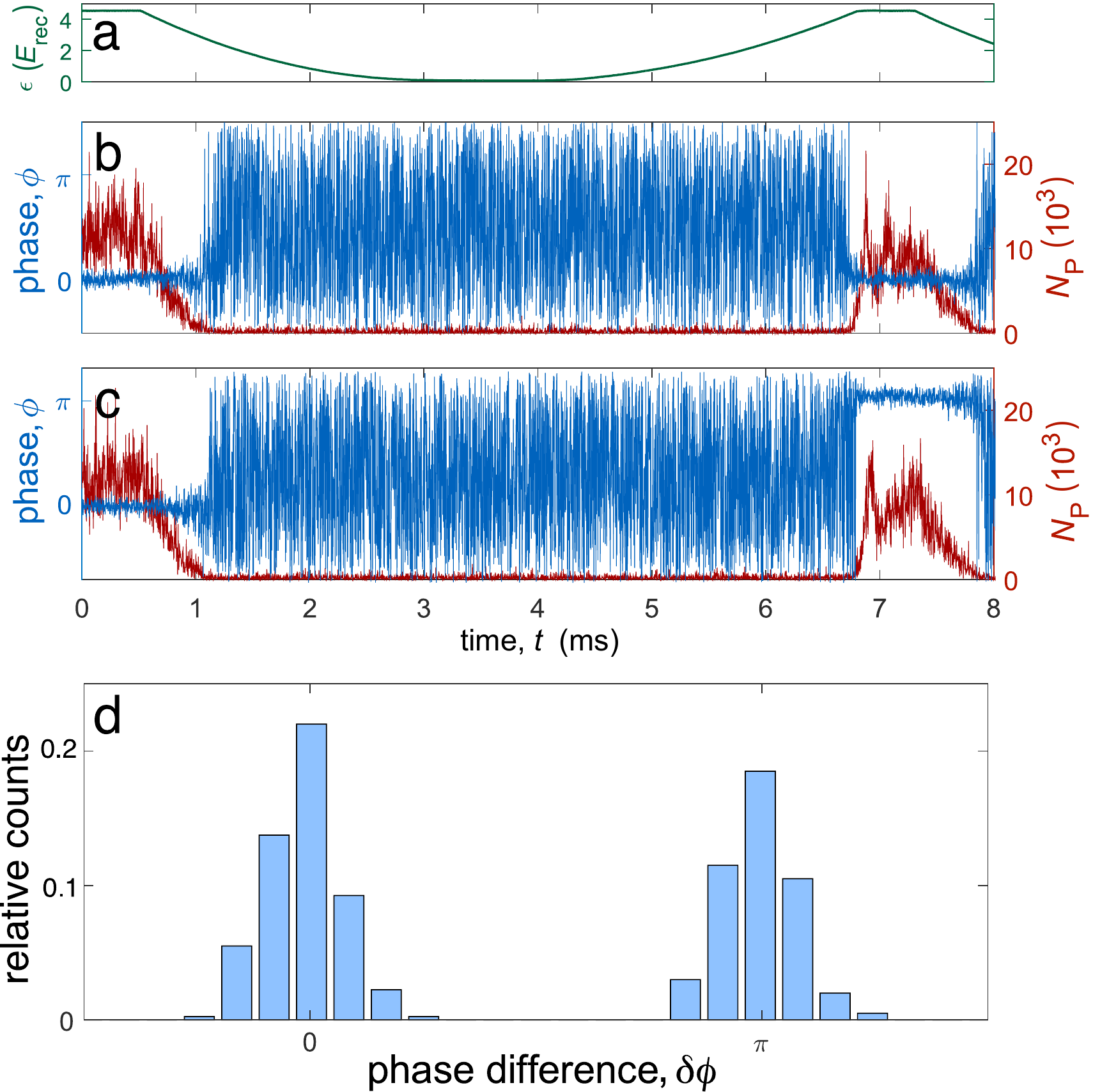}
\caption{{$\mathbb{Z}_2$ symmetry breaking in space.} {(a)} Pump protocol starting in the DW phase, tuning into the BEC phase and back to the DW phase. {(b),(c)} Relative phases $\phi$ (blue) and intracavity photon numbers $N_\mathrm{P}$ (red), measured by the heterodyne detector for single experimental runs, showing the two typical outcomes (b) $\delta \phi \approx 0$  and (c) $\delta \phi \approx \pi$. {(d)} Histogram of the phase difference $\delta \phi$ for 397 experimental runs.}
\label{sfig:1} 
\end{figure}

\section{$\mathbb{Z}_2$ symmetry breaking in space}
Due to optical path length drifts, we cannot compare the phase $\phi$ for DW realisations of different experimental runs. The stability of our balanced heterodyne detection is however sufficient to compare the phase for two subsequent DW realisations within the same experimental sequence applying the pump protocol in Fig.~\ref{sfig:1}(a). In a perfect system, the phase difference $\delta \phi$ between two subsequent realizations of the DW phase can take two values $\delta \phi = 0$ or $\delta \phi = \pi$ and does not depend on any system parameters. Since the underlying discrete symmetry breaks spontaneously, we expect equiprobable realisation of the two possible outcomes shown in Figs.~\ref{sfig:1}(b) and \ref{sfig:1}(c). The relative occurrence of $\delta \phi$ for 397 realisations is plotted in Fig.~\ref{sfig:1}(d) using a binning of $0.08~\pi$. The two maxima, corresponding to $\delta \phi = 0$ and $\delta \phi = \pi$, are clearly distinguishable and the ratio of all realisations where $\delta \phi \in[-\frac{\pi}{2},\frac{\pi}{2}]$ over $\delta \phi \in[\frac{\pi}{2},\frac{3}{2}\pi]$ is 1.15. This number is close to unity, which shows that the underlying spatial $\mathbb{Z}_2$ symmetry in our system is well established. 

\section{$\mathbb{Z}_2$ symmetry breaking in time}
\begin{figure}[tbp]
\centering
\includegraphics[width=0.9\columnwidth]{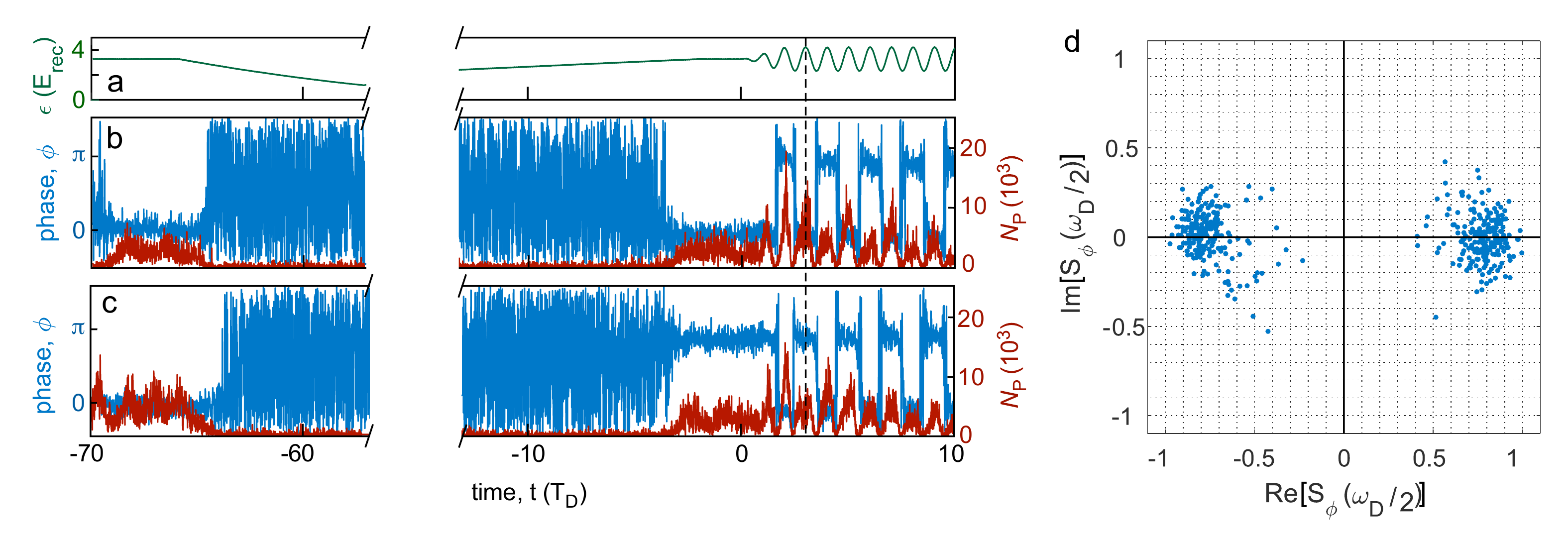}
\caption{Spontaneous breaking of the {$\mathbb{Z}_2$ time translation symmetry.} {(a)} Pump protocol starting in the DW phase, tuning into the BEC phase and back to the DW phase. After a waiting time of 0.5~ms the modulation strength $f_\mathrm{0}$ is linearly increased to $f_\mathrm{0}=0.3$. {(b),(c)} Relative phases $\phi$ (blue) and intracavity photon numbers $N_\mathrm{P}$ (red), measured by the heterodyne detector for single experimental runs showing the two typical outcomes $\delta \phi = 0$ (b) or $\delta \phi = \pi$ (c). As a consequence, also the time-phase difference between the subharmonic response and the modulated pump strength is constrained to the values zero and $\pi$. This is seen by observing the relative phase $\delta \phi$ at the time $t_{\textrm{max}}$, where the modulated pump strength acquires a maximum, indicated by the vertical black dashed line at $t \approx 3 T_D$. {(d)} Normalised Fourier component $S_\mathrm{\phi}(\omega_\mathrm{D}/2)$ of the relative phase $\delta \phi(t)$ rescaled by its maximum for 423 experimental runs.}
\label{sfig:2} 
\end{figure}

In this section we show how the $\mathbb{Z}_2$ symmetry breaking associated with the DW phase leads to a spontaneous breaking of the discrete $\mathbb{Z}_2$ time translation symmetry associated with the modulated pump strength. Again, the stability of the phase reference of our heterodyne detection system is not sufficient to compare the phases $\phi$ for different experimental runs. Therefore we follow a similar procedure as in Fig.~\ref{sfig:1}, entering the DW phase twice within the same experimental run. After entering the DW phase for the second time, we start to modulate the pump strength in the same way as in Fig.~1 of the main text. The applied pump protocol is presented in Fig.~\ref{sfig:2}(a). As discussed in Fig.~\ref{sfig:1}(d), the phase difference $\delta \phi$ between two subsequent realizations of the DW phase can take two values $\delta \phi = 0$ or $\delta \phi = \pi$ and does not depend on any system parameters. As a consequence, the time-phase difference between the observed subharmonic time-crystal oscillation and the oscillation of the pump strength becomes constrained to the possible values zero and $\pi$. This is seen by evaluating the relative phase $\delta \phi$ at the time $t_{\textrm{max}}$, where the modulated pump strength acquires a maximum, indicated by the vertical black line in Fig.~\ref{sfig:2}(a) at $t= 3 T_D$. Since the underlying discrete symmetry breaks spontaneously, we expect equiprobable realisation of the two possible outcomes shown in Figs.~\ref{sfig:2}(b) and \ref{sfig:2}(c). The normalised complex value of the Fourier spectrum at the subharmonic frequency, $S_\mathrm{\phi}(\omega_\mathrm{D}/2)$ of the relative phase $\delta \phi(t)$, rescaled by its maximum for 423 experimental runs, is shown in \ref{sfig:2}(d). The ratio between the occurrences with $Re[S_\phi(\omega_\mathrm{D}/2)]<0$ over the events with $Re[S_\phi(\omega_\mathrm{D}/2)]>0$ is 1.05, which shows that the discrete time translation symmetry associated with the modulation is well established.

\section{Theoretical model}
In the frame rotating at the pump frequency $\omega_\mathrm{P} = 2 \pi / \lambda_\mathrm{P}$, the Hamiltonian for the system reads\cite{Ritsch2013,Nagy:2008hk}
\begin{equation}\label{eq:2ham}
\hat{H} = \hat{H}_\mathrm{C} + \hat{H}_\mathrm{A} + \hat{H}_\mathrm{AA}  +\hat{H}_\mathrm{AC}.
\end{equation}
In Eq.~\eqref{eq:2ham}, the Hamiltonian for the cavity with a single mode function $\mathrm{cos}(kz)$ is 
\begin{equation}
 \hat{H}_\mathrm{C} = -\hbar\delta_{\mathrm{C}} \hat{a}^{\dagger}\hat{a},
\end{equation}
where  $\hat{a}$ ($\hat{a}^{\dagger}$) is the cavity mode annihilation (creation) operator. The single-particle Hamiltonian for the atoms is given by
\begin{equation}
\hat{H}_\mathrm{A} =\int dy dz \hat{\Psi}^{\dagger}(y,z)\left[-\frac{\hbar^2}{2m}\nabla^2 +  \epsilon\, \mathrm{cos}^2(ky) \right]\hat{\Psi}(y,z),
\end{equation}
where $m$ is the mass of an atom and $\hat{\Psi}(y,z)$ is the atomic field operator.  The short-range collisional interaction between the atoms is captured by the Hamiltonian
\begin{equation}
\hat{H}_\mathrm{AA} =U_\mathrm{a} \int dy dz \hat{\Psi}^{\dagger}(y,z)\hat{\Psi}^{\dagger}(y,z)\hat{\Psi}(y,z)\hat{\Psi}(y,z).
\end{equation}
The effective 2D interaction strength is $U_\mathrm{a} = \sqrt{2\pi}a_s \hbar^2/m\ell_x$, where $a_s$ is the $s$-wave scattering length and $\ell_x$ is the harmonic oscillator length in the $x$ direction. The Hamiltonian for the light-matter interaction reads
\begin{align}
\hat{H}_\mathrm{AC} =\hbar U_0 \int dy &dz \hat{\Psi}^{\dagger}(y,z)\biggl[ \mathrm{cos}^2(kz)a^{\dagger} a \\ \nonumber
&+ \alpha_{\mathrm{P}} \left( a + a^{\dagger} \right) \mathrm{cos}(ky) \mathrm{cos}(kz) \biggr]\hat{\Psi}(y,z),
\end{align}
where $\alpha_{\mathrm{P}} \equiv \sqrt{\epsilon/\hbar|U_0|}$ is the unitless amplitude of the pump field. The dynamics of the system follows from the Heisenberg-Langevin equations,
\begin{align}
\frac{\partial}{\partial t} \hat{\Psi} &= \frac{i}{\hbar}[\hat{H}, \hat{\Psi}] \\
\frac{\partial}{\partial t} \hat{a} &= \frac{i}{\hbar}[\hat{H}, \hat{a}] - \kappa \hat{a} + \xi,
\end{align}
where the stochastic noise term $\xi$ satisfies $\langle \xi^*(t)\xi(t') \rangle = \kappa \delta(t-t')$. We simulate the dynamics in the semiclassical limit by transforming $\hat{\Psi}$ and $\hat{a}$ into classical fields according to the truncated Wigner approximation (TWA) method \cite{Polkovnikov2010,Blakie2008,Carusotto2013}. 

\begin{figure}[!htbp]
\centering
\includegraphics[width=0.4\columnwidth]{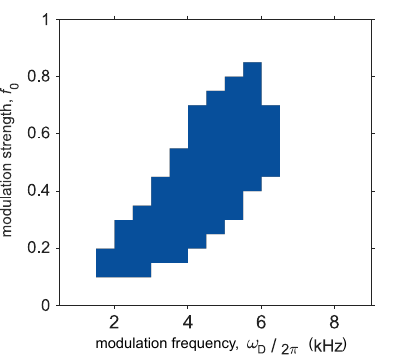}
\caption{{Mean-field stability region of the DTC.} Blue area denotes the region in the $(f_0,\omega_{\mathrm{D}})$-plane where a stable period-doubling response exists within the mean-field model in the absence of short-range interaction.}
\label{sfig:3} 
\end{figure}

The TWA is a semiclassical phase space method that goes beyond mean-field theory and can be utilised to test the robustness of time crystals against quantum and stochastic noise due to the dissipative cavity \cite{Cosme2019,Kessler2020}. We numerically integrate the resulting equations of motion for an ensemble of $10^3$ initial conditions, which sample the initial quantum noise in the fields and the stochastic noise due to the dissipative cavity. In our simulations, apart from $\epsilon_0$ chosen as $\epsilon_0 = 1.03~\epsilon_{\mathrm{cr}}$, where $\epsilon_{\mathrm{cr}}$ is the critical pump strength for the BEC-DW phase transition, we use the same parameters and protocol for the pump strength as in the experiment. In the comparison of calculations of $N_{\mathrm{P}}(t)$ and $C(t)$ for variable collisional interaction strengths $E_a$ in Fig.~4 in the main text, we adjust the pump strengths such that the number of intracavity photons in the DW phase is fixed to the same value.
\\ \\
\section{Mean-field phase diagram}
\begin{figure}[!htbp]
\centering
\includegraphics[width=0.7\columnwidth]{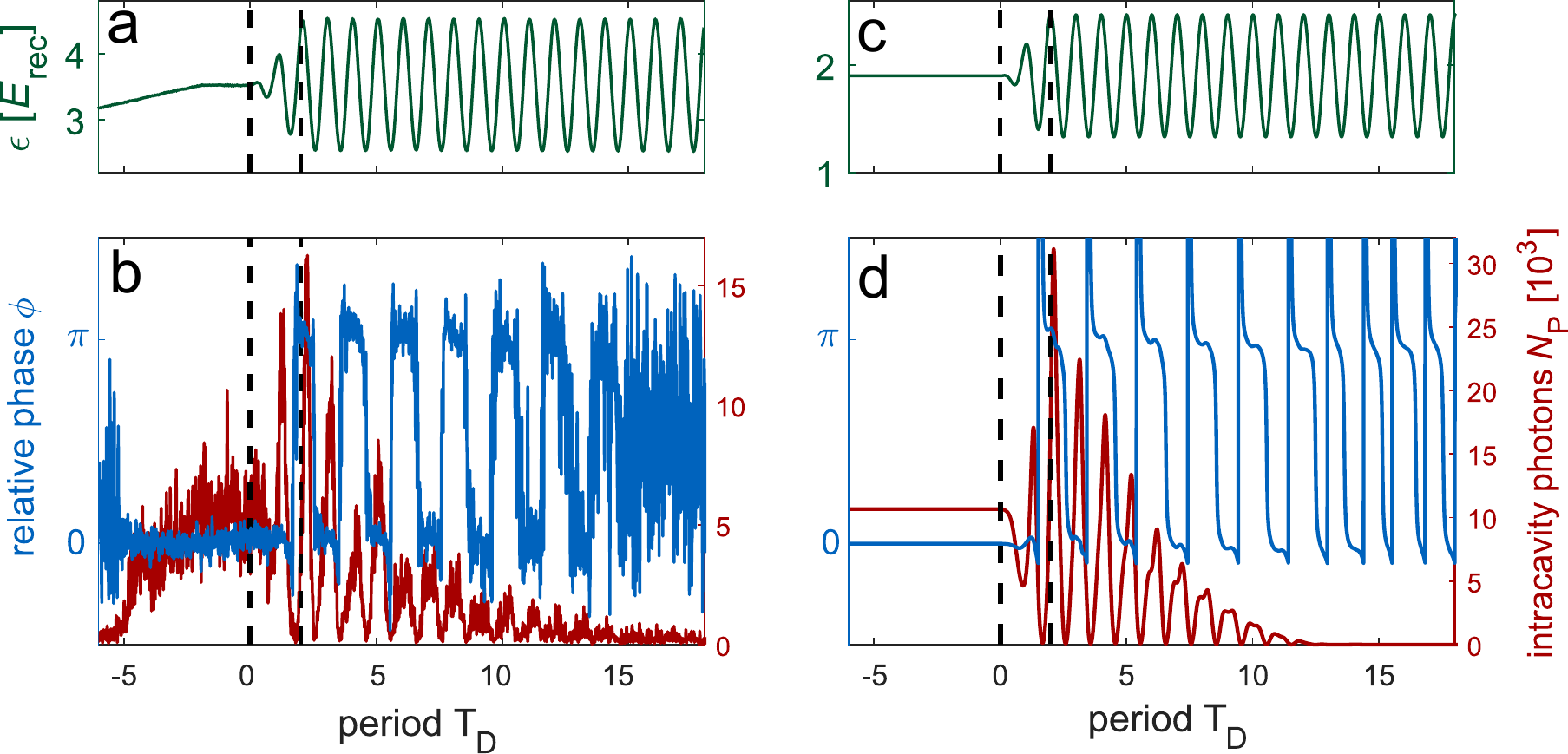}
\caption{{Comparison of experimental data to mean-field simulations including phenomenological atom loss.} {(a)} Time sequence for the pump with modulation strength $f_\mathrm{0}=$~0.3 and modulation period $T_\mathrm{D} = 0.25\,$ms. In the time interval delimited by dashed lines, $f_\mathrm{0}$ is linearly ramped from zero to its desired value. {(b)} The corresponding response of the intracavity photon number $N_\mathrm{P}$ (red) and the relative phase $\phi$ between the pump and the cavity light field (blue).  {(c),(d)} Corresponding mean-field simulations including atom loss.}
\label{sfig:4} 
\end{figure}

In order to obtain a rough orientation with regard to the system parameters suitable for the appearance of a dissipative time crystal (DTC) phase, we construct a dynamical phase diagram in the clean mean-field limit, wherein the mean-field breaking short-range interaction is neglected. In particular, we seek period-doubling solutions, which are stable for at least 40 modulation cycles. As depicted in Fig.~\ref{sfig:3}, modulation frequencies in the range $\omega_{\mathrm{D}}\in 2\pi \times [2,8]~\mathrm{kHz}$ provide an island with a stable DTC phase. This is consistent with the experimental results in Fig.~2(g) in the main text. Fig.~\ref{sfig:4}(a) and \ref{sfig:4}(b) show single shot measurements of the evolution of the intracavity photon number $N_\mathrm{P}$ (red) and the relative phase $\phi$ between the pump and the cavity light field (blue). In Fig.~\ref{sfig:4}(c) and \ref{sfig:4}(d),  the corresponding mean-field simulations including phenomenological atom loss are presented.

\section{Theoretical results with Temporal Disorder}
\begin{figure}[!htbp]
\centering
\includegraphics[width=0.6\columnwidth]{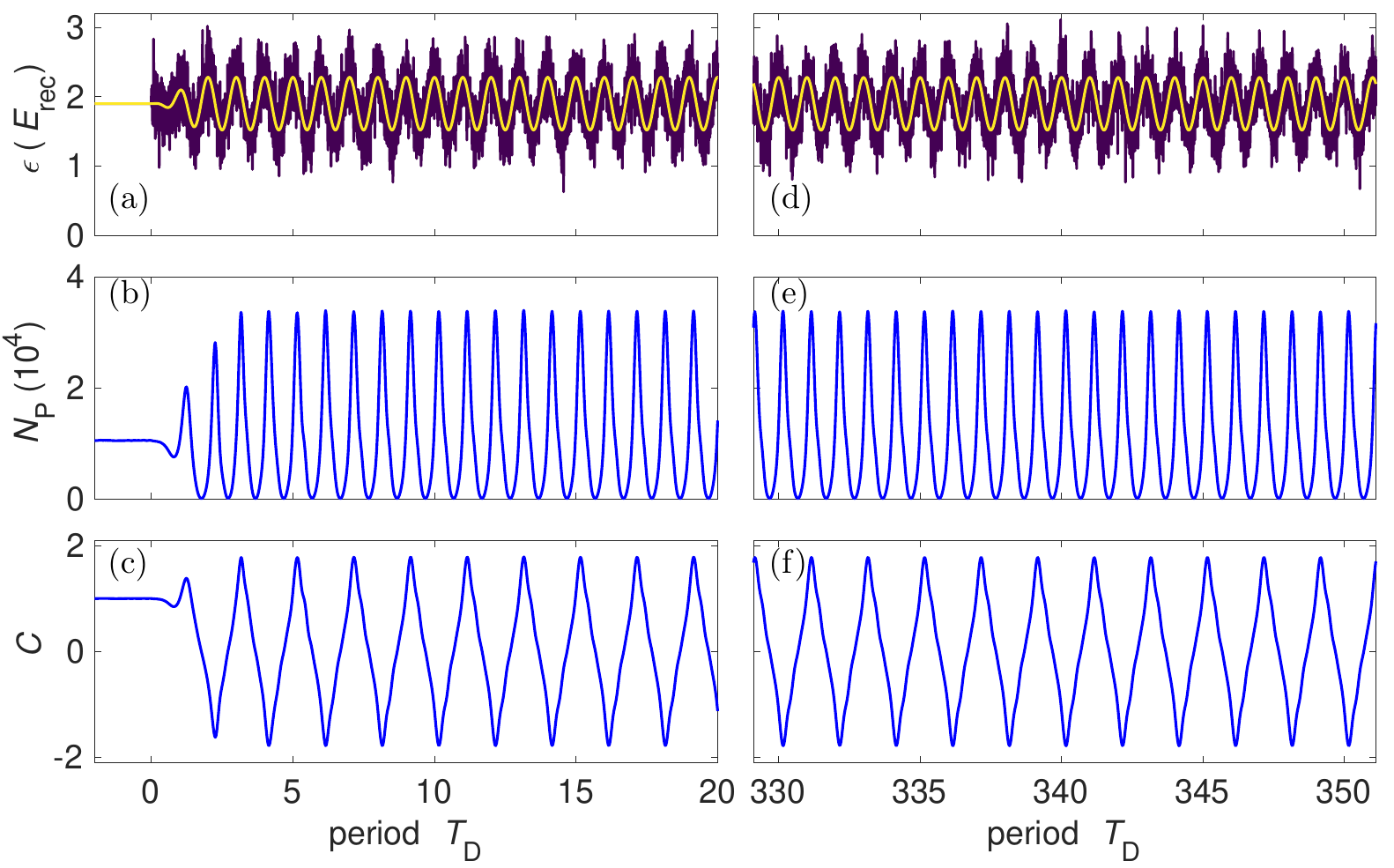}
\caption{Numerical results from TWA for noisy drive. {(a)-(c)} Short-time and (d)-(f) long-time dynamics.  (a),(d) Single realization of the disordered drive [dark] and the clean periodic drive [light].  TWA results for the (b),(e) intracavity photon number and (c),(f) non-equal time correlation. The modulation strength is $f_\mathrm{0}=$~0.3 and modulation period is $T_\mathrm{D} = 0.25\,$ms. }
\label{sfig:5} 
\end{figure}
In Fig.~\ref{sfig:5}, we present the results of our TWA simulations for a noisy drive.  Specifically,  we add a Gaussian white noise onto the pump strength signal.  An exemplary trace of the noisy drive is shown in Figs~\ref{sfig:5}(a) and \ref{sfig:5}(b). Note, however, that the noise in our numerical results shown here is band-limited to 0.025 GHz, which is set by the integration step of our stochastic differential equation solver. In contrast, the noise in the experiment is band-limited to 50 kHz. This explains the appearance of a more intermittent noise in the pump signal when Fig.~\ref{sfig:5} is compared to Fig. 3 in the main text. Similar to the experiment, we quantify the noise strength by $n \equiv \sum_{\omega}  |\mathcal{E}_\mathrm{noisy}(\omega)|/\sum_{\omega}|\mathcal{E}_\mathrm{clean}(\omega)|$,  where $\mathcal{E}_\mathrm{noisy}$ ($\mathcal{E}_\mathrm{clean}$) is the Fourier spectrum of the pump in the presence (absence) of white noise. The noise strength used in Fig.~\ref{sfig:5} is $n=2.0$. For this relatively weak temporal disorder, which breaks the discrete time translation symmetry imposed by the drive, our TWA results suggest that the system still exhibits long-lived period-doubling without any sign of decay after $\sim 350$ driving cycles. This corroborates the robustness of the DTC against temporal perturbations as explored experimentally in the main text.

\section{Mapping to the Dicke model}

The period-doubling instability of the DTC can be understood using a simple albeit incomplete description according to the mapping of the full atom-cavity Hamiltonian onto the Dicke model via the Schwinger-boson representation. Using the Holstein-Primakoff representation in the thermodynamic limit of $N \to \infty$, the collective spin in the Dicke model can be transformed back into bosons, leading to a coupled oscillator system, where the coupling strength is periodically driven. This coupled oscillator Hamiltonian can then be diagonalised to obtain a Hamiltonian for the lower and upper polaritonic states, where their respective frequencies are parametrically driven. Thus, driving at twice the lower polariton frequency leads to an exponential instability, which translates to a period-doubling response in the full atom-cavity model due to the presence of dissipation and the nonlinearity of the cavity-mediated interaction between the atoms.

\bibliography{references}